\documentclass{elsart}

\usepackage{natbib, graphicx}

\usepackage{amssymb}
\journal{Physics Letters A}
\begin{document}

\begin{frontmatter}



\title{Truncated L\'evy distributions in an inelastic gas}


\author[brenig,lambi]{R. Lambiotte},
\corauth[cor]{Corresponding author.} 
\ead{rlambiot@ulb.ac.be}
\author[brenig]{L. Brenig}
\ead{lbrenig@ulb.ac.be}

\address[brenig]{Physique Statistique, Plasmas et Optique Non-lin\'eaire, Universit\'e Libre de Bruxelles, Campus Plaine, Boulevard du Triomphe, Code Postal 231, 1050 Bruxelles, Belgium}
\address[lambi]{SUPRAS,  Sart-Tilman, Universit\'e de Li\`ege, B5, B-4000 Li\`ege, Belgium}

\begin{abstract}
We study a one-dimensional model for granular gases, the so-called Inelastic Maxwell Model. We show theoretically the existence of stationary solutions of the unforced case, that are characterized by an infinite average energy per particle. Moreover, we verify the quasi-stationarity of these states by performing numerical simulations with a finite number of particles, thereby highlighting  truncated L\'evy distributions for the velocities.

\end{abstract}


\end{frontmatter}

\section{Introduction}

High energy tails are generic for the velocity distributions of fluidized granular media, i.e. low density systems of macroscopic particles performing inelastic interactions.  Contrary to the specific shape of the tail which depends on the details of the grains, overpopulation seems to be a general feature of inelastic systems,  which expresses their deep non-equilibrium nature. Recently, classes of simplified models, the so-called Inelastic Maxwell Models (IMM) have been introduced. These models derive from a mathematical simplification of the non-linear Boltzmann equation for inelastic hard spheres, assuming that the collision rate may be replaced by an average rate, independent of the relative velocity of the colliding particles. This model introduces mathematical simplifications which may help to highlight the origin of the anomalous velocity distributions and serve as a simple tool to study far-from-equilibrium systems.
Quoting M.H. Ernst, one can consider that  "what harmonic oscillators are for quantum mechanics and dumb-bells for polymer physics, is what elastic and inelastic Maxwell models are for kinetic theory," (ref.\cite{ernst2}).

Usually, people focus on two kinds of  solutions of the inelastic Boltzmann equation, namely  scaling solutions describing the approach towards total rest state and heated stationary solutions.
In the following, however, we focus on another class of solutions of inelastic gases, which are stationary states of the unforced case. Before going further, let us point the recent exhaustive study of Ben-Naim and Machta (ref.\cite{naim2}), that  focuses independently on the same issue. Our analysis is done in the context of the one-dimensional IMM, where we show that stable L\'evy distributions play a central role. 
The reasons for studying these stationary L\'evy velocity distributions are multiple. First, they correspond to  exact solutions for an inelastic gas, and this fact is very rare. By discussing the physical relevance of these states, whose mean energy diverges, we show that there exists physical solutions which are very close to these stationary states and which may be considered {\em quasi stationary}. This paper is also motivated by the revived interest regarding generalized thermodynamical equilibrium theories. Indeed, long tail distributions have been found in diverse fields of science and are the object of a growing number of researches.  From the point of view of statistical physics, several attempts have been made in order to show physical mechanisms underlying these anomalous behaviours, such as 
 Tsallis's non-extensive entropy-based statistics (ref.\cite{tsallis}).The study of the above L\'evy distributions could help to build a thermodynamic-like foundation to  processes such as anomalous diffusion, fractal dynamics... This problem, which has first been questioned by Montroll and Schlesinger (ref.~\cite{montroll}),  has recently lead to study  simple kinetic models  in order to show  equilibrium-like features  of L\'evy distributions (ref.~\cite{barkai1}, \cite{zanette}). In this work, we will also discuss the range of validity of such L\'evy states, and show that they are limited to the context of Maxwell-like kinetic models.

\section{L\'evy velocity distribution}

The one-dimensional IMM has been first introduced by Ben-Naim and Krapivsky (ref.\cite{bennaim1}), and derives from  a simplification of the Boltzmann equation for inelastic hard rods.  The collision rule between grains 1 and 2 reads:

\begin{equation}
v_{1}^{'} = v_{1} - \frac{(1+ \alpha )}{2 \alpha} v_{12}  ~~~~
v_{2}^{'} = v_{2} + \frac{(1+ \alpha )}{2 \alpha} v_{12}
\end{equation} 
where the prime velocities are the pre-collisional velocities. The inelasticity is accounted through the so-called normal restitution coefficient $\alpha$ restricted to the interval [0:1].
Standard procedures, such as molecular chaos hypothesis, lead to the one-dimensional inelastic Boltzmann equation. The IMM follows after the replacement of the collision rate by an average rate and a rescaling of the time scale:

\begin{equation}
 \frac{\partial f(v_{1};t)  }{\partial t}  +f(v_{1};t) = \int_{-\infty}^{\infty}  dv_2 ~   (\frac{1}{\alpha}) 
f(v^{'}_{1};t) f(v_2^{'};t)   
\label{imm}
\end{equation} 
 where one has furthermore assumed that the system is and remain homogeneous in the course of time.
 Finite energy solutions of (\ref{imm}) have a natural tendency to reach the asymptotic solution $\delta(v)$, expressing the natural cooling down due to the dissipative interactions.
These solutions have been proven to exhibit multiscaling (ref.\cite{bennaim1}), i.e the system is characterized by an infinite set of independent cooling rates. Moreover, as shown by Baldassarri et al. (ref.\cite{balda1}), multiscaling leads to a power law  scaling velocity solution. 

It is well known that the Maxwell like collision operator of (2) is very similar to a convolution process (ref.\cite{ernst3}). This property is clear after using the Bobylev Fourier transform method to the kinetic equation:
 
\begin{equation}
 \frac{\partial \phi(k;t)   }{\partial t}  + \phi(k;t)  =  \phi((\frac{1+\alpha}{2}) k;t) \phi((\frac{\alpha-1}{2}) k;t)  
\label{boby}
\end{equation}
where  $\phi(k;t)$ is the characteristic function of the velocity distribution, defined by: $\phi(k;t) = \frac{1}{2 \pi} \int dv f(v, t) e^{i k v}$. This expression  reveals the mechanisms which dominate the long time dynamics of the kinetic model. Indeed,  the evolution of $\phi(k^{*};t)$ is determined by the values of $\phi(k;t)$ for $|k|\leq |k^{*}|$ (this comes directly from the inequalities $|(\frac{1+\alpha}{2}) k^{*}| \leq |k^{*}|$ and $|(\frac{1-\alpha}{2}) k^{*}| \leq |k^{*}|$). Consequently, as time increases, $\phi(k^{*};t)$ is determined by the initial condition $\phi(k;0)$ for smaller and smaller values of $|k|$ and, in the long time limit, the characteristic functions are dominated by their initial behaviour at $k\rightarrow0$.  If the initial energy diverges, and if the small k development writes $\phi(k)=1+a(t) |k|$, one easily shows that  $
 \partial_t a(t) =0
 $, i.e. the quantity $a$ describing the small k singularity of $\phi(k)$ is a constant of motion. In that case,  the stationary solution of (\ref{boby})  is the L\'evy distribution:
\begin{equation}
\label{levy}
\phi(k) = e^{-a |k|}
\end{equation}
which, in velocity space rewrites:
\begin{equation}
\label{levyVelocity}
f(v) = \frac{a}{\pi} \frac{1}{(a^2+  v^2)}
\end{equation}
Let us stress that this class of states is characterized 
by an infinite energy, a condition which is a priori necessary in order to make compatible the inelasticity of the interactions and the stationarity of the velocity distribution. Moreover, analogies with the Central-Limit theorem suggest that  (\ref{levyVelocity}) is the asymptotic stationary solution for a large class of initial distributions, thereby drawing a parallel with the Maxwell-Boltzmann equilibrium distribution (ref.\cite{lambi}). One should also remark that  the quantity  $a$ also shares some properties with an equilibrium temperature, such as the equipartition property in the case of mixtures (ref.\cite{barkai1}). 

\section{Physical relevance and simulations of L\'evy distributions}

\begin{figure}
\includegraphics[angle=-90,width=5.0in]{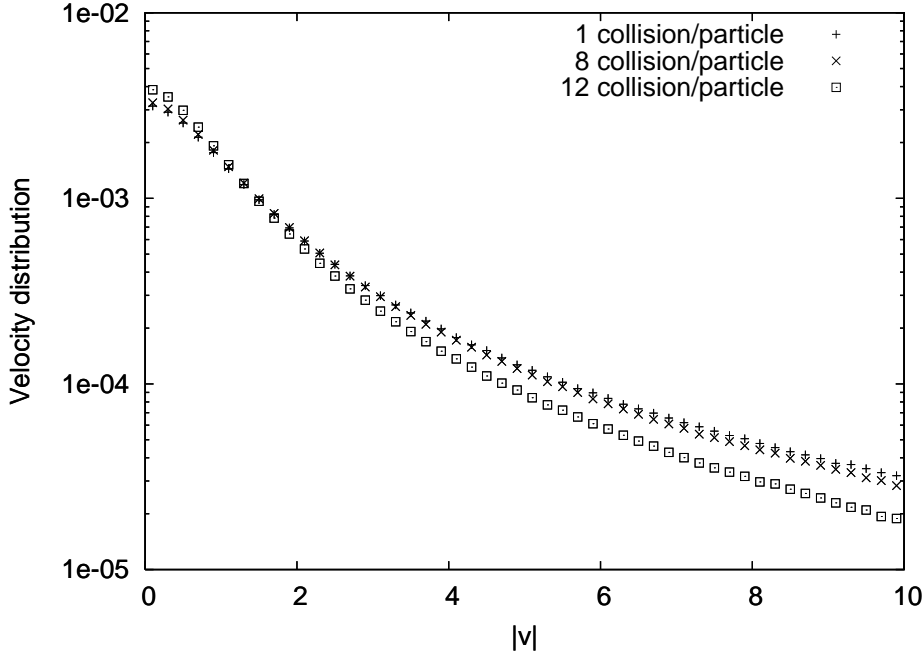}

\caption{ \label{dist}  Evolution of the core of the velocity distribution in log scale. The cutoff length of the L\'evy distribution is 10000. The inelasticity is $\alpha=0.5$ and the number of particles 10000000. We plot the velocity distribution after 1, 8 and 12 collisions per particle.}
\end{figure}

\begin{figure}
\includegraphics[angle=-90,width=5.0in]{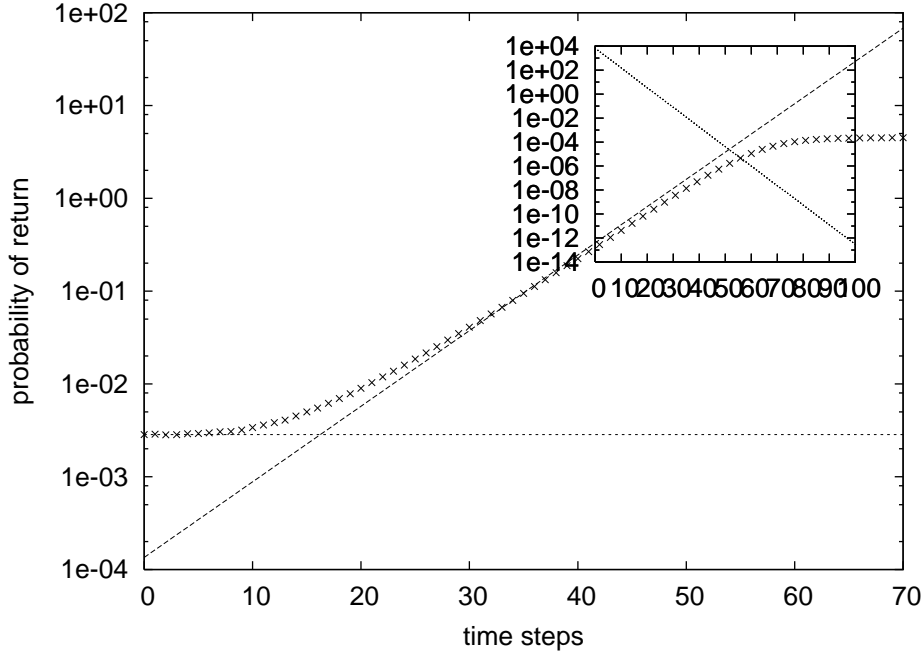}

\caption{ \label{transition} Evolution of the probability of return $f(0;t)$ for the same system as in figure (\ref{dist}). The lines correspond to the theoretical predictions of the two regimes. The asymptotic saturation is due to the finite size of the measurement cells.}
\end{figure}

In this section, we would like to discuss the physical relevance of the stationary L\'evy velocity distributions derived in the previous section. At first sight, the only physical stationary solution should be the total rest state, $\phi(k)=1$ ($e^{-a |k|}$, with a=0), because an infinite average  energy for the velocity distribution  sounds unphysical. It is nonetheless interesting  to point out that, despite this infinite average behaviour, each realization of the statistical process is finite, i.e. each particle carries a finite but unbounded energy. 
Moreover, physical systems  are composed by a finite number of constituents. Combining these two remarks, one expects that the average energy of a physical system is finite and decreases because of dissipation according to Haff law: $T(t) = T_0 ~ e^{- \frac{1- \alpha^2}{2} t}$, even if its velocities are initially L\'evy distributed.   In order to understand this paradox, we have  performed computer simulations of the Inelastic Maxwell Model. This can be achieved by using the usual stochastic interpretation of the Boltzmann equation (ref.\cite{bird1}). To do so, we initially distribute velocities according to a Lorentz distribution, and check that  the energy of the system decreases like the exponential law prescribed by theory. The velocity distribution function  is also shown to converge towards the Dirac delta distribution in the long time limit, as expected. In contrast, a careful study of the evolution of the core of the velocity distribution shows that it remains stationary during a non-negligible time interval (fig.~\ref{dist}). This property, together with the finiteness of the number of particles, allows one to make a link between the IMM and truncated L\'evy distributions (TLD). This family of distributions has been introduced by Mantegna and Stanley (ref.~\cite{mont}, \cite{schlesinger}) in order to justify the use of L\'evy-like distributions in finite-size systems.  They are defined by:
 
 \begin{eqnarray}
f_{trunc}^{\beta}(x)  &=& f_{L}^{\beta}(x) ~~  if ~ |x|< l \cr
f_{trunc}^{\beta}(x)  &=& 0 ~~  if ~ |x|> l
\end{eqnarray}
 where l is the cutoff length and $f_L^\beta$ is the L\'evy distribution of order $\beta$ defined to be the inverse Fourier transform of $e^{- a |k|^\beta}$. 
 In a simulation of the IMM, the cut-off length is naturally imposed by the upper boundary value of the energy of the particles.  In order to evaluate the stationarity of the L\'evy core distribution, we have measured the evolution of $f(0;t)$. Asymptotically, this variable should behave like $f(0;t) \sim e^{\frac{1- \alpha^2}{4} t}$, which is predicted from the scaling solution of Baldassari (ref.\cite{balda1}). In the simulations, we observe the transition between two linear regimes in logarithmic scale (fig.~\ref{transition}): an initial regime during which $f(0;t)$ is constant which corresponds to the quasi stationary regime of the core of the velocity distribution, and a linearly increasing regime, which comes from the energy decrease in the system. Amazingly, while $f(0;t)$ increased of only ten percent during ten mean collision times, the energy decreased by a factor 100 during the same time period.
 
\section{Conclusions}

As a conclusion, we would like to discuss the range of validity of the stationary velocity distributions described above. Indeed,
the Inelastic Maxwell Model studied in this paper is exceptional because of its strong similarities with random walk theory, which do not apply in general to more complex kinetic models.Therefore, it is important to recall a qualitative difference between elastic and inelastic systems, namely the fact that the asymptotic solutions of inelastic gases depend on their modeling while elastic systems always converge towards the Maxwell-Boltzmann distribution. This universality relies on the conservation of energy and on the time symmetric dynamics of elastic systems. Altogether, these properties imply the emergence of detailed balance at equilibrium. In contrast,
it is straightforward to prove that  stationary L\'evy distributions do not exhibit detailed balance (ref.\cite{lambi}), thereby showing the dependence of the stationary solution on the collisional kernel. This result confirms that L\'evy distributions are  associated to a summation of random variables, while the Gaussian also arises from more general  principles. 
 
Nevertheless, Maxwell models constitute an ideal framework, due to their mathematical simplicity, in order to study non-trivial features of the Boltzmann equation, such as non-equipartition of energy in inelastic mixtures, or  infinite energy solutions of the elastic case (ref.~\cite{boby1}). Moreover, they lead to  a natural derivation of fractional Fokker-Planck equations (ref.~\cite{barkai1}), and to the emergence of L\'evy distributions, whose interest in physics is therefore enhanced. 
 In this paper, we also discussed the physical relevance of such infinite mean energy solutions by performing computer simulations of the model. Given the finiteness of the number of particles in the system, we showed that these solutions are equivalent to truncated L\'evy velocity distributions, i.e. to a finite mean energy solution. It is important to stress that such a solution emerges from the dynamics itself, and not from an ad hoc assumption for the truncature. Moreover, computer simulations indicate that, while the total energy decays exponentially fast, as predicted by theory, the core of the distribution remains stationary during an arbitrary long period of time. Finally, we would like to point the recent work by Ben-Naim and Machta (ref.\cite{naim2}), who propose an experimental realization of  infinite energy steady states by 
injecting energy at large velocities. This experiment should open new perspectives in kinetic theory, and motivate detailed studies of this family of solutions.

We would like to thank  Dr. Figueiredo for fruitful discussions.
R.L. would also like to thank especially Dr. Ausloos and Dr. Mareschal for invitations at the Universit\'e de Li\`ege and at CECAM respectively.  This work has been done partly thanks to the financial support of FRIA, and ARC.




\end{document}